\documentclass[aps,prapplied,reprint,superscriptaddress]{revtex4-2}
\usepackage{amsmath}
\usepackage{graphicx}
\usepackage{bbm}
\usepackage[colorlinks,linkcolor=blue,anchorcolor=blue,urlcolor=blue,citecolor=blue]{hyperref}
\usepackage{graphicx}
\usepackage{dcolumn}
\usepackage{bm}
\usepackage[utf8]{inputenc}
\usepackage[T1]{fontenc}
\usepackage{etoolbox}
\usepackage{amssymb}
\usepackage{yhmath}
\makeatletter

\begin{document}

\title{Enhancement of photon blockade via topological edge states}

\author{Jun Li}
\email[]{jli\_phys@tongji.edu.cn}
\affiliation{MOE Key Laboratory of Advanced Micro-Structured Materials, School of Physics Science 
	and Engineering, Tongji University, Shanghai, 200092, China}
\affiliation{Department of Physics and Astronomy, University of Manitoba, Winnipeg R3T 2N2, Canada}

\author{C.-M. Hu}
\email[]{hu@physics.umanitoba.ca}
\affiliation{Department of Physics and Astronomy, University of Manitoba, Winnipeg R3T 2N2, Canada}

\author{Yaping Yang}
\email[]{yang\_yaping@tongji.edu.cn}
\affiliation{MOE Key Laboratory of Advanced Micro-Structured Materials, School of Physics Science 
	and Engineering, Tongji University, Shanghai, 200092, China}

\date{\today}

\begin{abstract} 
	Quantum technologies, holding the promise of exponentially superior performance than their classical 
	counterparts for certain tasks, have consistently encountered challenges, including instability in 
	quantum light sources, quantum decoherence and vulnerability to losses that topological photonics 
	happens to adeptly address. Here, we theoretically put forth a quantum Su-Schrieffer-Heeger-type 
	chain designed to greatly enhance single-photon blockade (single-PB) effect with topological 
	protection. By designing the deliberate coupling strengths, the quantum-level lattices take the form 
	of a one-dimensional array with a topological edge state in single-excitation space and a 
	two-dimensional square breathing lattice with topological corner states in two-excitation space, 
	resulting in enhanced single-photon excitation and the suppression of two-photon transitions. 
	Therefore the second-order correlation function is diminished by up to two orders of magnitude at 
	the cavity resonance frequency, accompanied by stronger brightness.Furthermore, the PB effect is 
	robust to local perturbations in cavity-qubit coupling and qubit frequency, benefitting from 
	topological protection.
\end{abstract}

\maketitle

\section{Introduction}

Since 2005, topological photonics\cite{Lu2014, RevModPhys.91.015006}, inspired by topological 
band theory initially discovered in solid-state electron systems, was first proposed in photonic 
crystals \cite{PhysRevLett.100.013904,Wang2009} and so far has been extensively investigated 
across various platforms, including waveguides \cite{nature12066}, cavities \cite{nphys2063,
	nphoton.2012.236}, metamaterials \cite{Li:18}, optomechanics \cite{Schmidt:15}, ultra-cold atoms 
\cite{RevModPhys.91.015005}, rf circuits  \cite{PhysRevLett.114.173902, PhysRevX.5.021031} and 
Fock-state lattices \cite{nwaa196, science.ade6219}. Further, Optical systems also offer 
unprecedented opportunities to promote the exploration of novel topological physics and actualize 
exceptional topological phenomena owing to their remarkable flexibility, diversity and unique 
effects, e.g., non-Hermitian topology \cite{RevModPhys.93.015005}, Floquet topological insulators 
\cite{nature12066,5.0063247}, topological photonics in synthetic dimensions \cite{Lustig:21}
and nonlinear topological photonics \cite{1.5142397}.

In the context of bulk-edge correspondence, topological edge states (TESs) in open boundary 
conditions are naturally robust against local defects and disorders, attributable to the globally 
defined topological invariants. This characteristic enables a diverse array of devices and 
applications for addressing pervasive environmental effects by employing topological photonics, 
such as wireless power transfer \cite{PhysRevApplied.15.014009,ZHANG2021}, optical beam 
splitter \cite{lpr}, sensor \cite{Guo:21} and nonreciprocity \cite{5.0063247} in one dimensions 
based on Su–Schrieffer–Heeger (SSH) model \cite{PhysRevLett.42.1698} and topological laser 
\cite{science.aar4003,science.aar4005}, slow light \cite{Yoshimi:20,PhysRevLett.126.027403}, 
channel-drop filter \cite{1.3593027} and four-wave mixing \cite{sciadv.aaz3910} in two dimensions. 
In particular, quantum light, acting as carriers of quantum information and quantum computing, 
still encounter challenges related to energy dissipation and decoherence stemming from external 
environment and system disorder. The emergence of topological photonic states represents a 
captivating research avenue to tackle this challenge \cite{adom.202001739,8848487}. In recent 
years, some pioneering experiments have showcased the robust generation and routing of  
quantum states in topological photonic integrated platforms \cite{science.aaq0327,
	science.aau4296,Mittal2018,Wang2019}.

Photon blockade (PB) effect serves as a significant technique for generating quantum light 
sources by effectively suppressing certain photon-number excitations in nonlinear systems. 
There are two physical mechanisms on which PB relies, i.e., the conventional photon blockade 
(CPB) \cite{Birnbaum2005,PhysRevLett.118.133604} and the unconventional photon blockade 
(UPB) \cite{PhysRevLett.104.183601,PhysRevA.96.053810,PhysRevLett.121.043601}. Specifically, 
The CPB is achieved through eigenenergy-level anharmonicity (ELA) originating from strong 
nonlinearity. The latter is induced by quantum destructive interference (QDI) between two or more 
individual quantum transition pathways. Generally, in comparison to CPB, the UPB exhibits a higher 
degree of anti-bunching, but the brightness is relatively poor. Recently, photon anti-bunching and 
corresponding mean photon number are both substantially improved by combining the conventional 
ELA-induced and QDI-induced single-photon blockade (single-PB) in a two-qubit driven cavity 
quantum electrodynamics (QED) system with dipole-dipole interaction \cite{Zhu:21}. Interestingly, 
the UPB is proposed to be enhanced in a chain of coupled resonators with exponentially suppressed 
nonlinearity requirement \cite{PhysRevLett.127.240402}.

In this work, we propose a one-dimensional (1D) topological multi-qubits-one-cavity chain that can 
achieve single-PB with much smaller second-order correlation function accompanied by larger 
cavity mean photon number than Jaynes-Cummings (JC) model (single-qubit-cavity system) under 
same conditions \cite{Birnbaum2005}. The quantum topological chain comprises a linear optical 
cavity and a trivial dimer qubit lattice, established through the weaker coupling between the cavity 
and the boundary qubit. The quantum-level lattice (QLL) in single-excitation space forms as an SSH 
chain with odd sites and possesses a zero-energy TES localized at the end of single-photon excited 
state, thereby enabling a stronger intensity of single-photon excitation. Meanwhile, in two-excitation 
space,  the zero-energy dressed state is localized at the corner of two-qubit excited states with 
absence of a two-photon excited state distribution. Furthermore, the dressed states with near the 
zero-energy in the second manifold are the bulk states that hold quite weak distribution of two-photon 
excited state. So that the single-PB effect is significantly enhanced by the distinct TES distributions 
in the two manifolds. Moreover, the single-PB effect in the quantum topological lattices also 
demonstrates robustness against the local disturbances in the cavity-qubit coupling strength and 
the qubit frequency.

\section{Model and quantum-level lattices}

\begin{figure}
	\centering\includegraphics[width=5 cm]{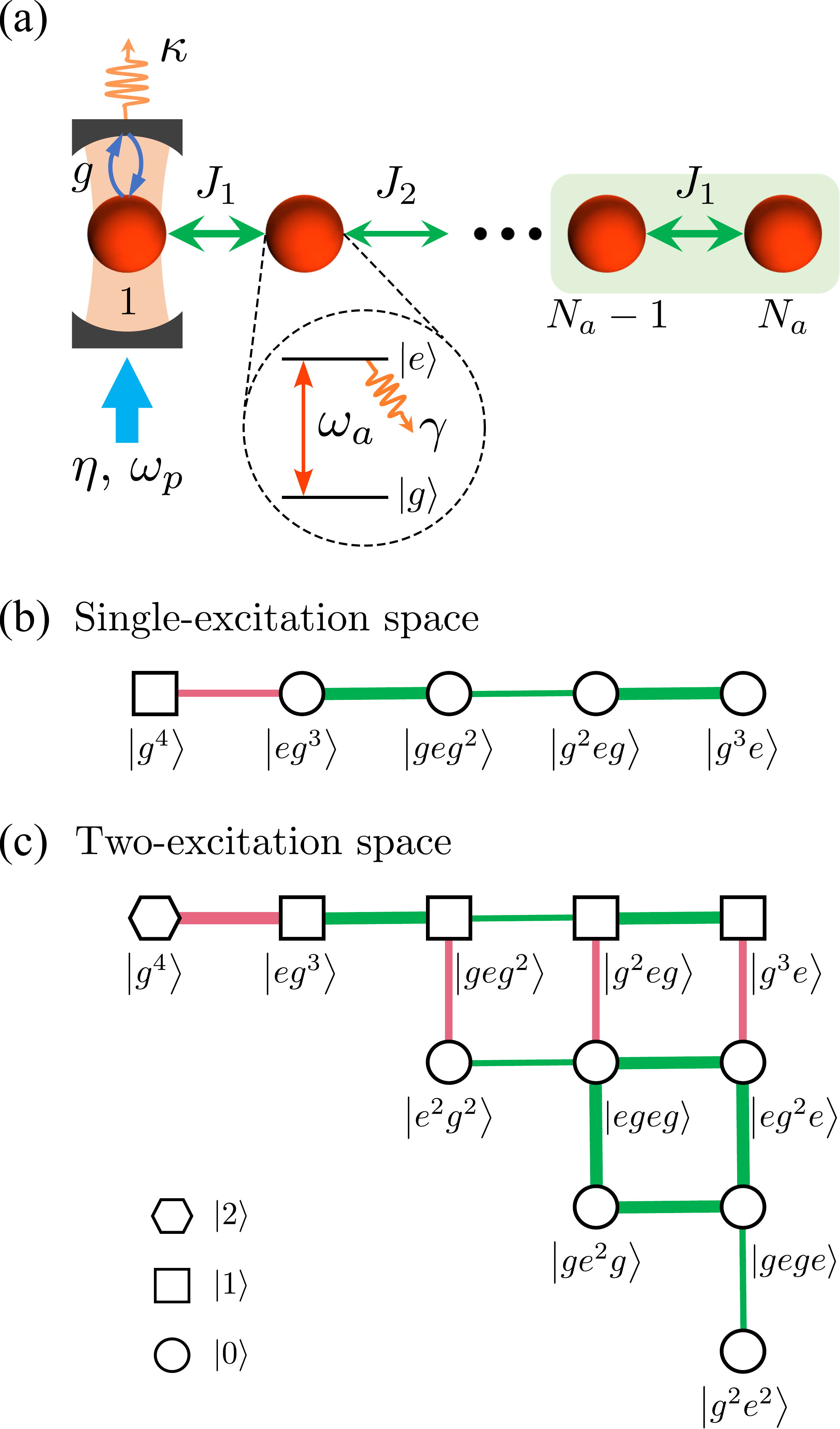}
	\caption{\label{Fig.1} (a) Schematic of considered multi-qubits-cavity topological chain characterized 
		by an optical cavity (resonant frequency $\omega_c$) coupled to the first site of a dimer two-level 
		lattice with the strength $g$. The dimer lattice, whose unit cell are marked by a green box, consists 
		of $N_{a}$ identical two-level qubits with the transition frequency $\omega_a$ between ground 
		state $\left|g\right\rangle$ and excited state $\left|e\right\rangle$. The intracell and intercell 
		hopping amplitudes of the lattice are $J_{1}$ and $J_{2}$, respectively. The cavity (qubit) decay 
		rate is $\kappa$ ($\gamma$). QLL in (b) single-excitation space and (c) 
		two-excitation space for $N_{a}=4$. The states labeling the lattice sites are the corresponding 
		positional distributions of the qubit states with the power of the exponent representing the number 
		of repetitions of adjacent qubit states. For clarity, we use circles, squares and hexagon in (b) and 
		(c) to denote the cavity photon states $\left| 0 \right\rangle$, $\left| 1 \right\rangle$ and $\left| 2
		\right\rangle$, respectively. The widths of the lines connecting neighboring states represent the 
		coupling strengths between them.  
	}
\end{figure}

As shown in Fig. \ref{Fig.1}(a), we consider a 1D quantum topological array composed of a single-mode 
bosonic cavity and $N_a$ ($N_a=2N$, $N$ is the number of unit cells) identical two-level systems 
(e.g., two-level qubits) arranged in a dimer chain (or SSH lattice) by coupling to their nearest neighbors 
with alternative hopping amplitudes $J_1$ and $J_2$. The optical cavity is coupled to one of the 
outermost qubit with the strength $g$ and coherently driven by a monochromatic pump field with 
the Rabi frequency $\eta$ and angular frequency $\omega_d$. Here, all the coupling strengths are 
much smaller than the transition frequency of qubits $\omega_a$ and the resonant frequency of cavity 
$\omega_c$. In rotating-wave approximation, the Hamiltonian of the driven multi-qubits-cavity system 
is $H=H_{a}+H_{c}+H_{i}+H_{d}$, where (setting $\hbar$ = 1)

\begin{equation} \label{eq:1}
	\begin{split}
		H_{a} & = \Delta_{a} \sum_{i=1}^{N_{a}}\sigma_{+}^{i} \sigma_{-}^{i}+J_{1}\sum_{i=1}^{N_{a}/2} 
		\left(\sigma_{+}^{i} \sigma_{-}^{i+1} + \mathrm {H.c.} \right)\\ 
		&+ J_{2}\sum_{i=1}^{N_{a}/2-1} \left(\sigma_{+}^{i+1} \sigma_{-}^{i+2} +\mathrm {H.c.} \right),\\
		H_{c} & = \Delta_{c} a^{\dagger} a, \quad H_{i} = g \left(a^{\dagger} \sigma_{-}^{1} + 
		\mathrm{H.c.}\right),\\
		H_{d} &=  \eta \left( a^{\dagger} + a\right),
	\end{split}
\end{equation}
where $\Delta_{a} = \omega_{a}-\omega_{d}$ is the qubit detuning with respect to the driving frequency
and $\Delta_{c} = \omega_{c}-\omega_{d}$ is the cavity-driving detuning. Here, we focus on the simplest 
scenario where the frequency of the cavity and the qubits are identical ($\omega_{a} = \omega_{c} = 
\omega_{0}$), i.e., $\Delta_{a}=\Delta_{c}=\Delta$. $\sigma_{+}^{i} = | e \rangle \langle g |_{i}$ 
($\sigma_{-}^{i}= |g\rangle \langle e |_{i}$)  is the raising (lowering) operator of the $i$th two-level system 
and $a^{\dagger}$ ($a$) is the creation (annihilation) operation of the intracavity  optical field.

In single-excitation space, the QLL of the system takes the form of a 1D finite SSH chain with an odd 
number of sites ($N_{a}+1$), which is identical to its classical lattice counterpart [see an example for 
$N_{a} = 4$ shown in Fig. \ref{Fig.1}(b)]. The lattice is expanded to two dimensions and comprises 
$N_{a}(N_{a} + 1)/2 + 1$ sites in two-excitation space  for $N_{a} \geqslant 4$. The quantum states 
form a square breathing lattice with triangle boundaries, constituting a two-order topological structure 
as illustrated in Fig. \ref{Fig.1}(c). In particular, due to the inherent nature of the annihilation operator 
$a \left| n \right\rangle = \sqrt{n} \left|n-1\right\rangle$, the coupling strength between the two-photon 
state $\left|g^{N_a},2\right\rangle$ and its nearest neighbor state $\left|g^{N_{a} - 1}e, 1\right\rangle$ 
is $\sqrt{2} g$. Additionally, the QLL in three-excitation space is represented as a three-dimensional 
cubic lattice with a triangular pyramid as the outline. 

In this study, we consider the condition where $g<J_1$ and $J_2<J_1$ to ensure that the single-photon 
excited state $\left|g^{N_a},1\right\rangle$ is the sole weaker coupling termination in single-excitation 
lattice. Therefore, it can be inferred that there is an exactly zero-energy state exponentially localized to 
the end with one-photon excited state. Furthermore, we add the constraint $J_1\ \leqslant \sqrt{2}g$ in 
order to prevent the two-photon state $\left| g^{N_{a}}, 2 \right\rangle$ from being a relatively weaker 
coupling corner in two-excitation lattice for $N_{a} \geqslant 4$. Instead the two-qubit excited state 
$\left|g^{N_a-2}e^2,0\right\rangle$ always serves as a weakly coupled corner. Thus, the topological 
zero-energy state in two-excitation space is predominantly distributed at the two-qubit excited states 
rather than the two-photon excited state. By combining the aforementioned two conditions, it is 
expected that single-PB effect will occur at the resonance frequency of the cavity. 

Based on the Born-Markov approximation, the dynamic density matrix $\rho$ of the entire system with 
dissipation is governed by the master equation

\begin{equation} \label{eq:2}
	\frac{\partial \rho}{\partial t}=-i[H, \rho]+\frac{\kappa}{2} \mathcal{L}_{\rho}[a]+\frac{\gamma}{2}
	\sum_{i=1}^{N_{a}} \mathcal{L}_{\rho}[\sigma_{-}^{i}],
\end{equation}
where $\mathcal{L}_{\rho}[a] = 2a\rho a^{\dagger}-a^{\dagger} a\rho-\rho a^{\dagger} a$ indicates 
the cavity leakage at rate $\kappa$ and $\mathcal{L}_{\rho}[\sigma_{-}^{i}] = 2 \sigma_{-}^{i} \rho \sigma_{+}^{i}-\sigma_{+}^{i}\sigma_{-}^{i}\rho-\rho \sigma_{+}^{i}\sigma_{-}^{i}$ denotes the 
spontaneous decay of the excited state of $i$th qubit at rate $\gamma$.

\section{Enhancement of photon blockade}

\begin{figure*}
	\centering\includegraphics[width=\linewidth]{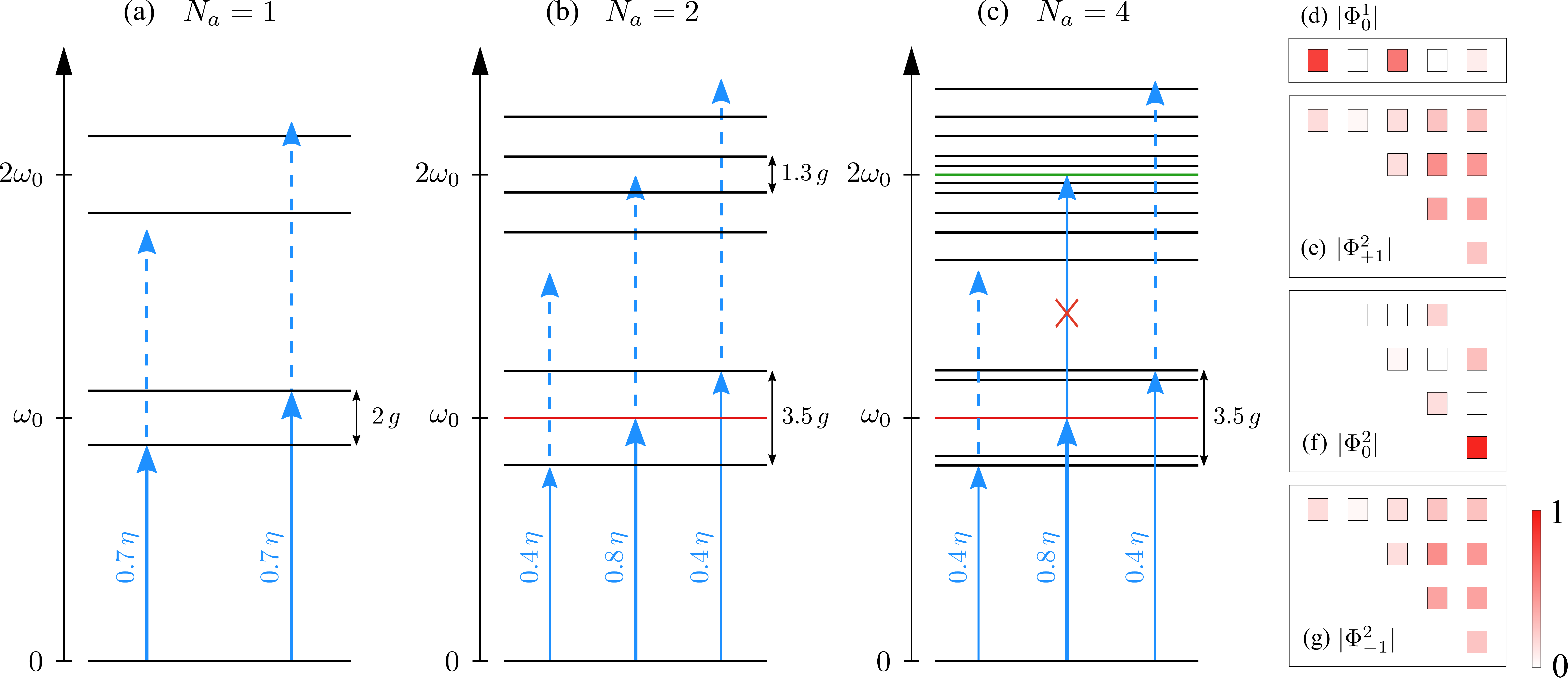}
	\caption{\label{Fig.2} The dressed state structures and the main transition pathways of considered 
		qubit-cavity coupled system with the number of qubits (a) $N_a=1$ (JC model), 
		(b) $N_a=2$ and (c) $N_a=4$. The levels labeled by red in (b) and (c) are the topological edge 
		modes in single-excitation space while the green level in (c) is the topological corner state in 
		two-excitation space. The single blue arrows in (a-c) denote the excitation pathways of photon 
		blockade (PB) where the dashed arrows indicate the ELA. The wavefunction distributions of 
		(d) the topological zero-energy states $|\Phi_{0}^{1}$ in single-excitation space, (f) the 
		zero-energy corner-localized state $\Phi_{0}^{2}$ and (e, g) the bulk states $\Phi_{\pm1}^{2}$ 
		in two-excitation space for $N_{a}=4$. The layout of sites in (d) and (e-g) is accordance with 
		that shown in Fig. \ref{Fig.1}(b) and \ref{Fig.1}(c), respectively. Here, we take the parameter 
		relations $J_1=\sqrt{2} g$ and $J_2=0.2 g$.}
\end{figure*}

In the absence of the driven field and system dissipation, we can numerically obtain the eigenvalues 
$E_{\pm i}^{n}$ and their corresponding eigenstates $\Phi_{\pm i}^{n}$, where the superscript $n \in 
[0,1,2, \cdots]$ indicates the order of the manifold and the subscript $\pm i$ denotes $i$th symmetric 
eigenstates with reference to zero-energy in each space. By setting the normalized hopping 
amplitudes $J_1/g=\sqrt{2}$ and $J_2/g=0.2$, we plot the corresponding eigenvalue structures of 
our topological array for the simpler cases of $N_{a} = 2$ in Fig \ref{Fig.2}(b) and $N_{a} = 4$ in Fig 
\ref{Fig.2}(c), respectively. As expected, in the first manifold, there always exists an exact zero-energy 
dressed state $\Phi_{0}^{1}=C_{1} \left|g^{N_{a}},1\right\rangle + C_{2} \left|eg^{N_{a-1}},0 \right\rangle 
+\cdots + C_{N_{a}+1} \left|g^{N_{a-1}}e,0 \right\rangle $ that is exponentially localized at the 
single-photon excited state $ \left|g^{N_{a}},1\right\rangle$ (as exemplified in Fig. \ref{Fig.2}(d) for 
$N_{a}=4$), with the normalized amplitudes $C_{1}>0.8$ and $C_{2x} = 0$ ($x\in [1, \cdots, N]$) as 
$N_{a}$ varies. In the second manifold, the zero-energy dressed level is absent for the simplest case 
of $N_{a} = 2$ as shown in Fig \ref{Fig.2}(b). For $N_{a} \geqslant 4$, there are $Na/2-1$ zero-energy 
states primarily distributed among the two-qubit excited states, without any distribution at the 
two-photon excited state (as shown in Fig. \ref{Fig.2}(f) for $N_a=4$). Moreover, for comparison, we 
include the dressed state diagram of coupled single-qubit-cavity system (JC model) in Fig \ref{Fig.2}(a), 
where there is no zero-energy state in any manifold.

The single-PB effect behaviors are quantified by the equal-time second-order correlation function 
$g^{(2)}(0)$, which can be calculated by
\begin{equation} \label{eq:3}
	g^{(2)}(0)=\frac{\left\langle a^{\dagger 2} a^2 \right\rangle} {\left\langle a^{\dagger}a\right\rangle^2} 
	= \frac{Tr\left(a^{\dagger 2} a^2 \rho\right)} {\left[Tr \left(a^{\dagger} a \right)\right]^2},
\end{equation}
where the density matrix $\rho$ is obtained by numerically solving Eq. (\ref{eq:2}) based on the QuTiP 
\cite{JOHANSSON2012} and the QuantumOptics.jl toolbox \cite{kramer2018}. 

\begin{figure*}
	\centering\includegraphics[width=\linewidth]{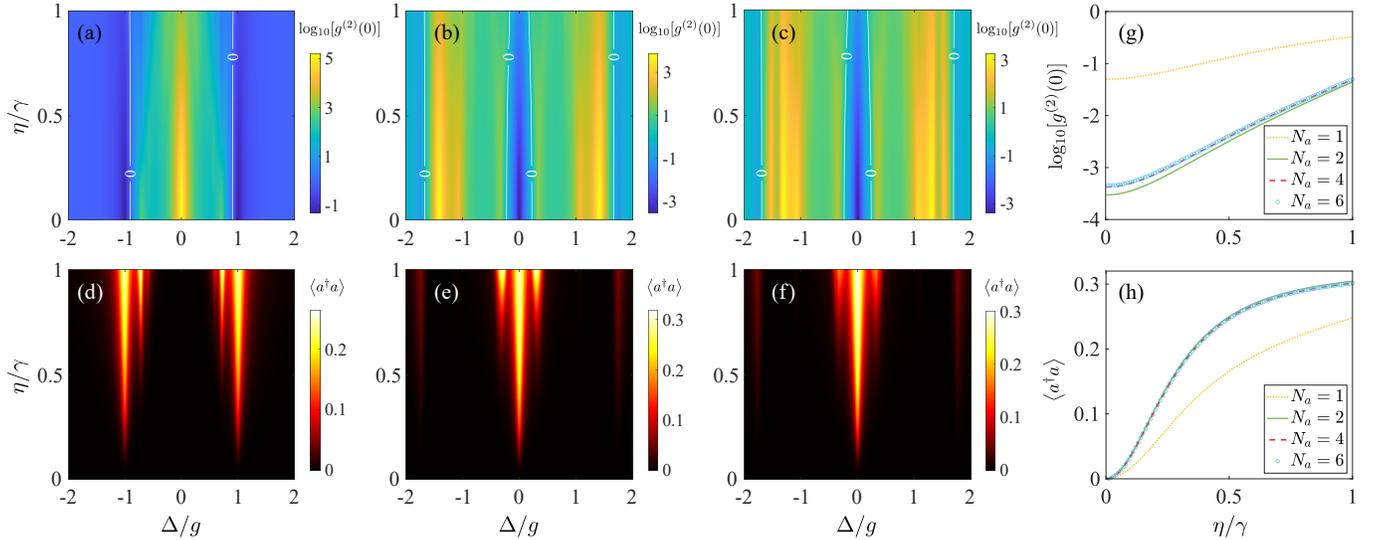}
	\caption{\label{Fig.3} (a-c) The equal-time second-order correlation function $\log_{10}[g^{(2)}(0)]$ 
		and (d-f) the corresponding mean photon number $\left\langle a^{\dagger}a\right\rangle$ mapping 
		on the plane of normalized detuning $\Delta/g$ and driven strength $\eta/\gamma$ for (a, d) $N_{a}
		= 1$, (b, e) $N_{a} = 2$ and (c, f) $N_{a} = 4$, respectively. The white curves in (a-c) denote the 
		contours of $g^{(2)}(0) = 1$ for Poissonian statistics. (g) The optimal second-order correlation 
		function and (h) the corresponding mean photon number versus the normalized pump field Rabi 
		frequency $\eta/\gamma$ for different $N_{a}$. The optimal condition are $\Delta/g = \pm 1$ for 
		$N_{a} = 1$ and $\Delta = 0$ for $N_{a}\geqslant 2$. Here, the system parameters are given by 
		$g=10\gamma$ and $\kappa=0.5 \gamma$. The other parameters are the same as those used in 
		Fig. \ref{Fig.2}.}
\end{figure*}

With the same normalized system parameters $g/\gamma=10$ and $\kappa/\gamma = 0.5$, figures 
\ref{Fig.3}(a-c) show the logarithmic plots of equal-time second-order correlation function $\log_{10}
[g^{(2)}(0)]$ as a function of the detuning $\Delta/\gamma$ and driven strength $\eta/\gamma$ for 
different numbers of qubits $N_{a}= 1$, 2 and 4, respectively. The corresponding mean photon number 
$\left\langle a^{\dagger}a\right\rangle$ are displayed in Figs. \ref{Fig.3}(d-f). In the single-qubit case, 
single-PB occurs around the detuning $\Delta=\pm g$ due to the single-photon resonance with 
non-resonant two-photon excitations as depicted in Fig. \ref{Fig.2}(a) \cite{Birnbaum2005,
	PhysRevLett.118.133604}. For our multi-qubits-cavity topological chain, as clearly shown in Figs. 
\ref{Fig.3}(b) and \ref{Fig.3}(c), there are three regimes of single-PB ($g^{(2)}(0) < 1$) with the 
detuning $\Delta = 0$ and $\Delta \approx \pm 1.8 g$ on the plane, which correspond to the 
single-photon excitation process from the ground state $ \left|g^{N_{a}},0 \right\rangle$ to the TES 
$\Phi_{0}^{1}$ and the bulk states $\Phi_{\pm N_{a}/2}^{1}$ of QLL in single-excitation space 
respectively. Particularly, the resonance-driven scenario shows a significantly smaller second-order 
correlation function, indicating strong single-photon antibunching. Meanwhile, higher mean photon 
number peak concomitantly appears due to the stronger single-photon excitation probability. 
Moreover, the frequency range for achieving single-PB around $\Delta = 0$ slightly expands as the 
number of qubits $N_{a}$ increases.

For the sake of comparison, we present the optimal second-order correlation function $\log_{10}
[g^{(2)}(0)]$ in the frequency domain against the normalized pump strength $\eta/\gamma$ for different 
numbers of qubits in Fig. \ref{Fig.3}(g), along with the corresponding mean photon number $\left\langle 
a^{\dagger}a\right\rangle$ in Fig. \ref{Fig.3}(h). Obviously, in comparison to the JC model ($N_{a}=1$), 
the multi-qubits-cavity lattices ($N_{a}\geqslant 2$) exhibit up to two orders of magnitude smaller 
$g^{(2)}(0)$, accompanied by larger mean photon number. For two-qubits case, the enhancement of 
single-PB arises from the greater two-photon excitation anharmonicity and the lower two-photon 
intensity of dressed states in the second manifold in contrast to the JC model. With an increasing 
number of qubits ($N_{a} \geqslant 4$), as depicted in Fig. 2(c), the number of dressed levels in the 
second manifold increases quadratically, resulting in smaller anharmonicity and the emergence of 
zero-energy dressed states $\Phi_{0}^{2}$. However, in the resonance-driven case, there is only a 
slight variation in the second-order correlation function and the mean photon number. On the one hand, 
the resonant transition $\Phi_{0}^{1} \rightarrow\Phi_{0}^{2}$ is strictly forbidden due to the lack of 
overlap between the wavefunction distributions of the zero-energy states in the two manifolds on the 
same qubit states, as shown in Figs \ref{Fig.2}(d) and \ref{Fig.2}(f). On the other hand, the dressed 
states with near-zero energy in the second manifold correspond to bulk states characterized by a weak 
intensity of two-photon state $\left| g^{N_{a}}, 2 \right\rangle$ [see Figs. \ref{Fig.2}(e) and \ref{Fig.2}(g)]. 
Further, the intensity decreases as the number of qubits $N_{a}$ increases, leading to the expansion of 
single-PB region around $\Delta = 0$.

\begin{figure}
	\centering\includegraphics[width=8 cm]{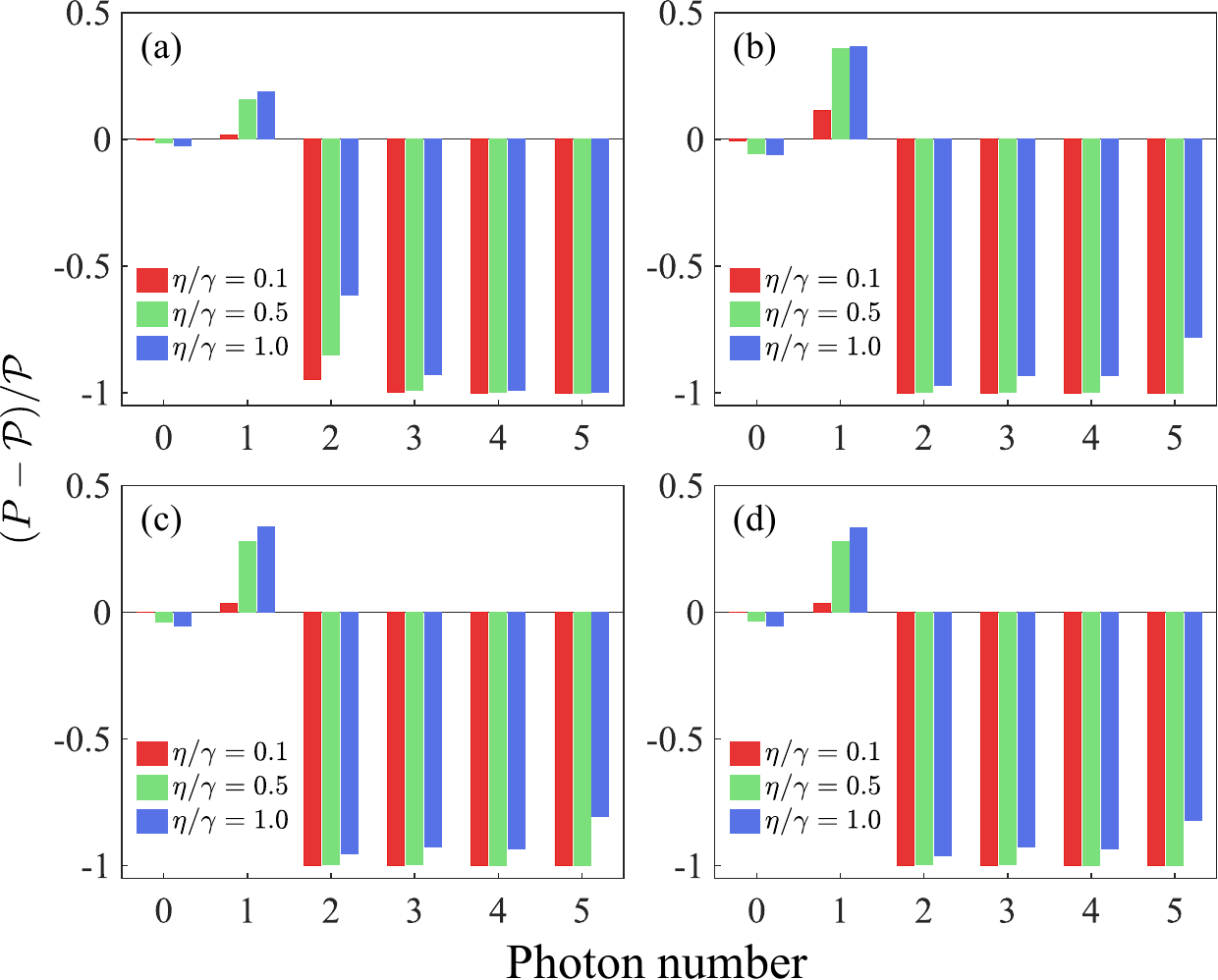}
	\caption{\label{Fig.4} The deviations of the photon distribution to the standard Poisson distribution 
		with the same mean photon number for (a) $N_{a}=1$, (b) $N_{a}=2$, (c) $N_{a}=4$ and (d) 
		$N_{a}=6$, respectively. The cavity-driving detuning is $\Delta = \pm g$ for $N_{a} = 1$ and 
		$\Delta = 0$ for $N_{a}\geqslant 2$. The other system parameters are the same as those used in 
		Fig. \ref{Fig.3}.
	}
\end{figure}

The enhancement of single-PB can be further confirmed by comparing the photon-number distribution 
of the cavity field $P(n)$ with the Poisson distribution $\mathcal{P}(n)=\left\langle a^{\dagger} a 
\right\rangle^{n} e^{-\left\langle a^{\dagger} a\right\rangle} / n!$. In Fig. \ref{Fig.4}, we show the relative 
deviations of photon number distribution with respect to the Poisson distribution with the same mean 
photon number, i.e., $\left[P(n)-\mathcal{P}(n)\right]/\mathcal{P}(n)$. In comparison to the JC model 
($N_a=1$) with the identical conditions, it is evident that $P(1)$ in the multi-qubits-cavity system 
($N_{a} \geqslant 2$) is more strongly enhanced while $P(n>1)$ are significantly suppressed, indicating 
the higher probability for detecting a single photon. Moreover, the probability distribution $P(1)$ is also 
greatly enhanced as the driven strength increases.

\section{Robustness of photon blockade}

\begin{figure*}
	\centering\includegraphics[width=\linewidth]{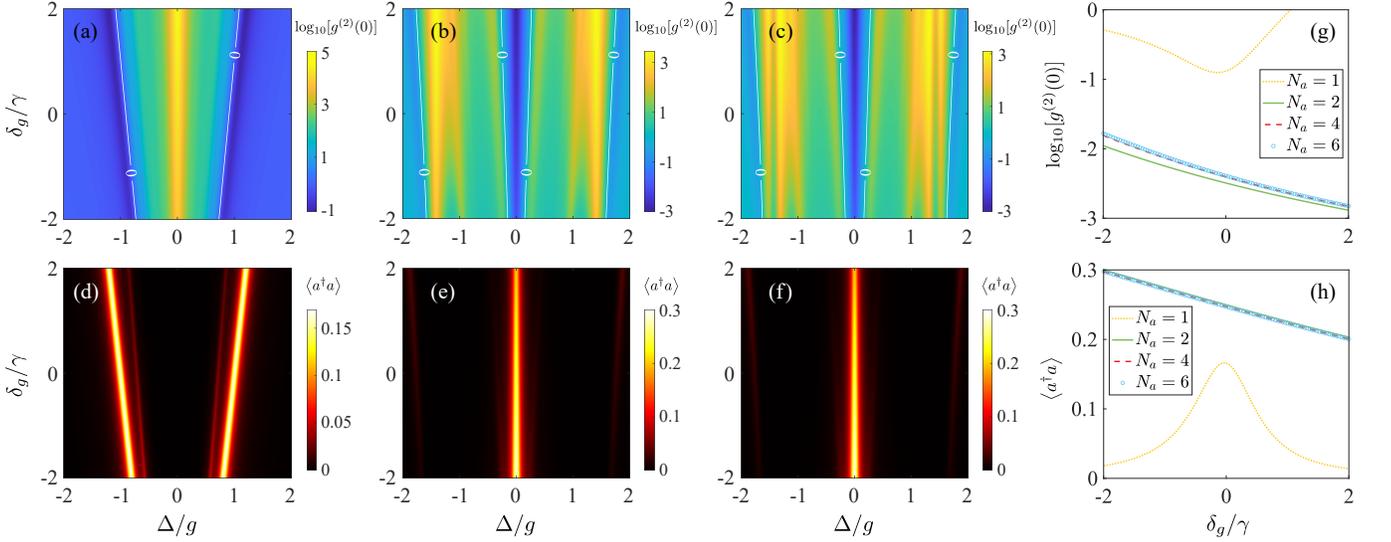}
	\caption{\label{Fig.5} The influences of fluctuation strength of cavity-qubit coupling $\delta_g$. (a-c) 
		Steady-state second-order correlation function $\log_{10}[g^{(2)}(0)]$ and (d-f) cavity mean 
		photon number $\left\langle a^{\dagger}a \right\rangle$ plotted as a function of the normalized 
		detuning $\Delta/g$ and fluctuation strength $\delta_g/\gamma$ with the driven strength 
		$\eta/\gamma = 0.5$ for (a, d) $N_{a}=1$, (b, e) $N_{a} = 2$ and (c, f) $N_{a} = 4$, respectively. 
		(g) The second-order correlations and (h) the corresponding mean photon number in relation to 
		the normalized fluctuation strength $\delta_g/\gamma$ with the detuning $\Delta = \pm g$ for 
		$N_{a} = 1$ and $\Delta = 0$ for $N_{a}\geqslant 2$. All other parameters remain the same as 
		those utilized in Fig. \ref{Fig.3}.
	}
\end{figure*}

The PB effect of our quantum topological array effectively incorporates and demonstrates the inherent 
topological protection property of the SSH chain that are insensitive to local perturbations. Here, we 
examine the robustness of the single-PB phenomenon in the presence of perturbations by considering 
specific examples of disturbances in the cavity-qubit coupling strength $\delta_g$ and the transition 
frequency of the first qubit $\delta_{\omega 1}$. In Fig. \ref{Fig.5}, by considering the cavity-qubit 
coupling strength $g'=g+\delta_g$, we present the evolution of the second-order correlation function 
$\log_{10}[g^{(2)}(0)]$ and the corresponding cavity mean photon number $\left\langle a^{\dagger}a 
\right\rangle$ versus the fluctuation strength of cavity-qubit coupling $\delta_g$ with $\eta/\gamma 
= 0.5$ for different $N_a$. Predictably, in the JC model, the optimal operating frequency for achieving 
single-PB experiences a linear shift in response to the fluctuation strength of the cavity-qubit coupling 
due to the linear coupling-dependent dressed levels in the first manifold. In contrast, in the case of  
multi-qubits-cavity arrays ($N_{a}\geqslant2$), the optimal operating frequency, which yields the 
smallest second-order correlation function $g^{(2)}(0)$ and the largest mean photon number 
$\left\langle a^{\dagger}a \right\rangle$, consistently remains at the resonance frequency of the cavity 
regardless of variations in the cavity-qubit coupling strength. For the resonance-driven condition, 
$g^{(2)}(0)$ and the mean photon number $\left\langle a^{\dagger}a \right\rangle$ decrease (increase) 
with increasing (decreasing) cavity-qubit coupling strength, owing to the heightened (depressed) 
two-photon excitation anharmonicity and single-photon excitation probability, respectively. As clearly 
shown in Figs. \ref{Fig.5}(g) and \ref{Fig.5}(h), compared with the single-qubit cavity system, the 
second-order correlation function and mean photon number of the quantum topological arrays are 
more robust to the cavity-qubit coupling fluctuation at a fixed operating frequency.

\begin{figure*}
	\centering\includegraphics[width=\linewidth]{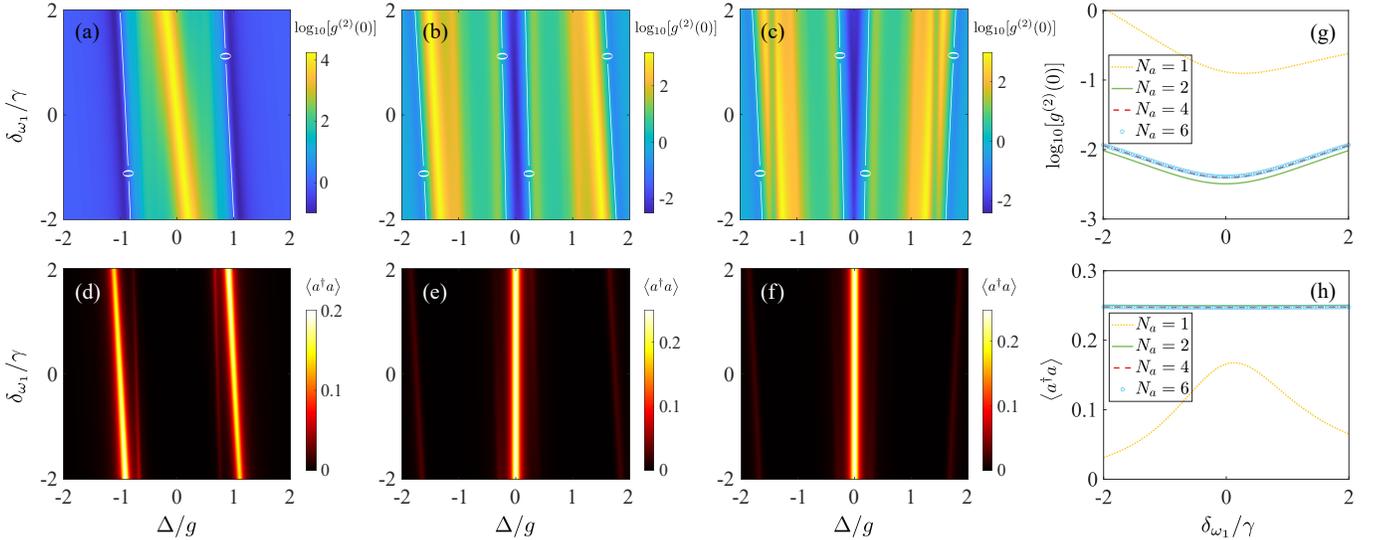}
	\caption{\label{Fig.6} The effects of first qubit's transition frequency shift $\delta_{\omega 1}$. Plots 
		of (a-c) $\log_{10}[g^{(2)}(0)]$ and (d-f) $\left\langle a^{\dagger}a\right\rangle$ as a function of 
		the normalized detuning $\Delta/g$ and frequency shift $\delta_{\omega 1} / \gamma$ with 
		$\eta/\gamma = 0.5$ for (a, d) $N_{a}=1$, (b, e) $N_{a}=2$ and (c, f) $N_{a}=4$, respectively. 
		Profiles of (g) the second-order correlation function and (h) the corresponding mean photon 
		number at the cavity-driving detuning $\Delta = g$ for $N_{a} = 1$ and $\Delta = 0$ for $N_{a}
		\geqslant 2$. The other system parameters are the same as those employed in Fig. \ref{Fig.3}.
	}
\end{figure*}

For the diagonal perturbation, we show the developments of correlation function and corresponding 
cavity mean photon number with the normalized frequency shift of the qubit coupled to cavity 
$\delta_{\omega 1}/\gamma$ as shown in Fig. \ref{Fig.5}. For single qubit condition, the optimal pump 
frequency to achieve PB with a brighter mean photon number is directly related to the qubit 
frequency, which is not conducive to manipulation in practical applications. Contrastingly, the 
multi-qubits-cavity chain maintains an almost constant mean photon number at the cavity resonance 
frequency, given that the frequency and distribution of TES in the single-excitation space are hardly 
affected by the transition frequency of the qubit. Simultaneously, in the two-excitation space, the 
distribution density of the zero-energy state on the two-photon excited state is no longer entirely 
zero with a nonzero qubit frequency shift. This leads to a gradual increase in the second-order 
correlation function with the increasing qubit frequency shift for the resonance-driven condition.

\section{Conclusion}
In summary, we have investigated an enhanced conventional photon blockade effect utilizing a 1D 
cascading coupled multi-qubits-one-cavity array based on the SSH model. One-sided localized 
topological edge states of the QLL in single-excitation space empower the system to exhibit a 
stronger intensity of single-photon excitation at the resonance frequency while two-photon 
excitation is effectively suppressed due to the forbidden resonance excitation and diffusely 
distributed bulk states in two-excitation space. The quantum topological chain exhibits a robust
single-PB effect, achieving up to a two-order magnitude improvement in the second-order 
correlation function yet enhancing the mean photon number simultaneously. Our work 
contributes a theoretical framework for the preparation of stable quantum light sources 
leveraging topological states.The implementation of our scheme can potentially be demonstrated 
in quantum systems such as coupled atom arrays cavity QED \cite{PhysRevLett.125.073601,
	PhysRevLett.125.073602} and circuit QED system \cite{PhysRevX.8.021058, science.abq5769}.

National Natural Science Foundation of China (12274326); National Key Research and Development 
Program of China (2021YFA1400600, 2021YFA1400602); China Scholarship Council (202106260079).
C.-M. Hu acknowledgments the support from NSERC Discovery Grants and NSERC Discovery 

\bibliography{ref}

\end{document}